\documentclass[aps,prb,twocolumn,groupedaddress]{revtex4-1}
\usepackage{hyperref}
\usepackage{epsfig}
\usepackage{amssymb,amsmath}
\usepackage{tikz}
\usetikzlibrary{decorations.pathreplacing}
\usepackage{standalone}
\usetikzlibrary{calc}
\usepackage{verbatim}
\usepackage{float}
\usepackage{lipsum,booktabs}
\usepackage{wrapfig,yfonts}
\usepackage{titlesec}
\usepackage{dcolumn}

\begin{document}

\title{Higher Order Bosonic Topological Phases in Spin Models}

\author{Oleg Dubinkin}
\email[]{olegd2@illinois.edu}
\author{Taylor L. Hughes}
\email[]{hughest@illinois.edu}

\affiliation{Department of Physics and Institute for Condensed Matter Theory, University of Illinois at Urbana-Champaign, Urbana, IL, 61801-3080, USA}

\date{\today}

\begin{abstract}
  We discuss an extension of higher order topological phases to include bosonic systems. We present two spin models for a second-order topological phase protected by a global $\mathbb{Z}_2\times\mathbb{Z}_2$ symmetry. One model is built from layers of an exactly solvable cluster model for a one-dimensional $\mathbb{Z}_2\times\mathbb{Z}_2$ topological phase, while the other is built from more conventional spin-couplings (XY or Heisenberg).  These models host gapped, but topological, edges, and protected corner modes that fall into a projective representation of the symmetry. Using Jordan-Wigner transformations we show that our models are both related to a bilayer of free Majorana fermions that form a fermionic second-order topological phase. We also discuss how our models can be extended to three-dimensions to form a third-order topological phase. 
\end{abstract}

\pacs{spin, SPT, Heisenberg, Quadrupole}

\maketitle

\section{Introduction}
As is well established by now, the nontrivial topology of the bulk band structure of topological insulators (TIs) and superconductors (TSCs), manifests itself through the presence of stable, gapless modes at the boundary of the system\cite{hasan2010}. Recently, this paradigm was extended to include a new class of higher-order topological phases (HOTPs)\cite{quadrupole,quadrupole2,Song2017,Langbehn2017,hoti2018,Trifunovic2018,Wang2018}. An $n$-th order HOTP defined on a $d$-dimensional lattice is a system with a gapped bulk which harbors gapped, topological phases on its co-dimension $1, 2,\ldots, n-1$ surfaces, and gapless protected modes realized on its $(d-n)$-dimensional boundaries. The first example of a 2nd order TI was the 2d topological quadrupole insulator (QTI) that has gapped, topological edges, and mid-gap corner modes\cite{quadrupole}. Before these works other models had appeared which host protected hinge or corner modes\cite{teo2013,Isobe15,benalcazar2014,Song16,benalcazar2018} and at least some of them exhibit higher-order topology.

In this article we propose bosonic counterparts of the QTI protected by a global $\mathbb{Z}_2\times\mathbb{Z}_2$ symmetry. We construct two models built from spin-$1/2$ objects that yield spin versions of a QTI. The first model we consider is built from layering and coupling pairs of $\mathbb{Z}_2\times\mathbb{Z}_2$ bosonic SPT chains with a cluster-model Hamiltonian. The resulting 2d model is exactly solvable and exhibits gapped edges which are in a 1d topological $\mathbb{Z}_2\times\mathbb{Z}_2$ SPT phase, and symmetry-fractionalized, gapless corner modes protected by the bulk $\mathbb{Z}_2\times\mathbb{Z}_2$ symmetry. The second model we consider has a more traditional spin-spin interaction and consists of dimerized XY (or Heisenberg) spin interactions on a square lattice with four spin-$1/2$ degrees of freedom per cell. We show that this model exhibits a similar higher order topological phase with dangling spin-$1/2$'s on the corners that are shared by the dimerized spin-$1/2$ SPT chains on the edges.
We then proceed to discuss how the latter model can be extended to 3d to form a 3rd order HOTP.

The paper is organized as follows: in Section \ref{hotpdef} we discuss how we plan to identify the bosonic HOTPs by analogy with the known fermionic cases. In Section \ref{z2z2}  we review the cluster-model Hamiltonian of a spin-$1/2$ $\mathbb{Z}_2\times\mathbb{Z}_2$ SPT phase, and then construct an exactly solvable 2d bosonic HOTP model from stacking and coupling the chains. We go on to exhibit the phenomenology of this model, and compare with the fermionic QTI. In Section \ref{xysec} we introduce a model with more conventional spin-spin couplings and exhibit its phenomenology. We then show how this model can be extended to three-dimensions to form a 3rd order HOTP. In Section \ref{conclusion} we discuss possible generalizations of these models and summarize our results. We include a series of appendices that discuss some details and extra justification for the arguments in the main article sections.

\section{Criterion for Identifying a Higher Order Topological Phase}\label{hotpdef}
\begin{figure}[h]
  \centering
  \includegraphics[width=0.5\textwidth]{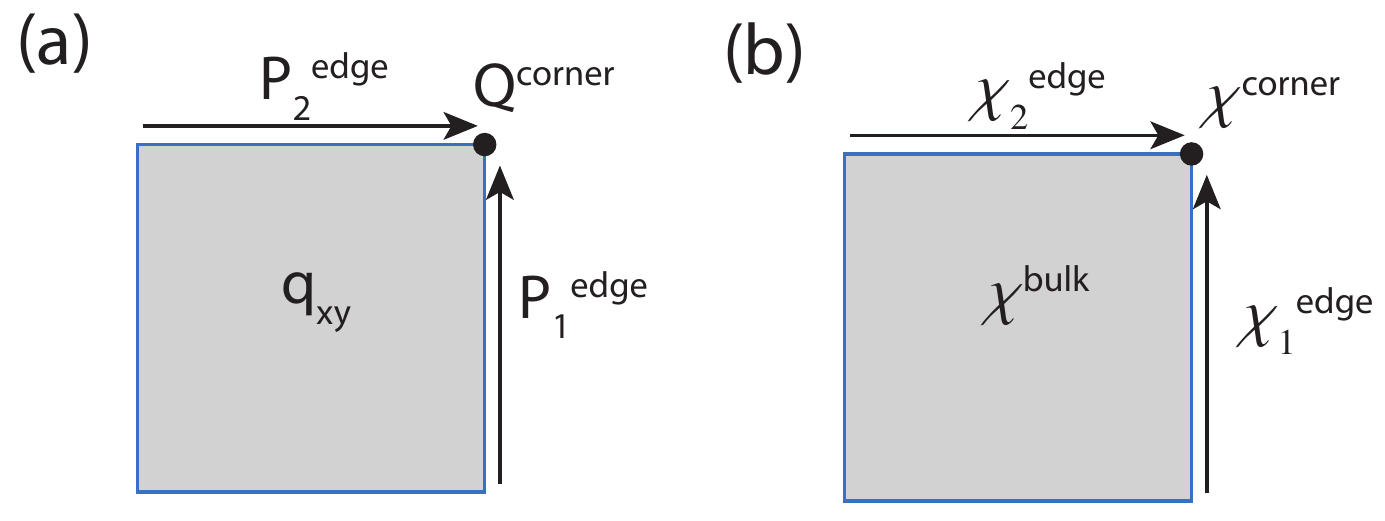}
  \caption{(a) Illustration of the bulk quadrupole moment, edge polarizations and corner charge (b) Illustration of bulk, edge, and corner topological indicators used to define a 2d HOTP.}
  \label{fig:hotpdef}
\end{figure}
At present there is no clear definition for a higher-order topological phase in a many-body system. The known classification tools used to identify fermionic HOTPs are built upon single-particle electronic structure, and have not been generalized to a many-body context yet. Unfortunately, one of the features that is most often associated to HOTPs, low-energy modes on higher co-dimension surfaces are neither necessary, nor sufficient indicators of higher-order topology. For example, 2d HOTPs often have topological modes on the corners of a sample, but there are other 2d topological phases with corner modes that are not higher order\cite{teo2013,benalcazar2014}, and there are HOTPs where the existence of corner modes depends on the details of the crystal termination\cite{quadrupole}.

Nevertheless, we can make progress by making an analogy to a known bulk characterization of fermionic HOTPs. Consider the 2d fermionic QTI with the quadrupole moment $q_{xy}\neq 0.$ The physical consequence of a bulk quadrupole moment is not an edge polarization or a corner charge, instead it is the relationship between the edge polarizations and the corner charge. If we take a square shaped sample and consider the charge on one of the corners, then the edge polarizations of the edges that intersect at that corner and the corner charge itself (all of which are quantized to take values of either $0$ or $1/2$) satisfy (see Fig. \ref{fig:hotpdef}(a))
\begin{equation}
Q^{corner}-P_{1}^{edge}-P_{2}^{edge}=-q_{xy}.
\end{equation} This relationship is crucial as it eliminates the contributions of any surface effects (e.g., edge reconstruction, changes in edge/corner termination) and allows  one to define the quadrupole moment in terms of the corner and edge properties. This relation captures the physical distinction between a system with edge polarization due to a bulk quadrupole moment and a system with edge polarization due solely to boundary effects. In the former case, i.e., the one of interest, the corner charge is 
\emph{shared} by the two edges, and in the latter case both edges contribute independently to the corner charge.

We will use this definition in our study of bosonic HOTPs. The analogous quantity to the corner charge in our systems will be the presence or absence of corner modes that form a projective representation of the symmetry group, just as one identifies 1d symmetry protected topological phases (SPTs) by the projective representations of their end modes. We will only be considering $\mathbb{Z}_2\times\mathbb{Z}_2$ symmetries so let us indicate the corner topology by the $\mathbb{Z}_2$ quantity $\chi^{corner}.$ The analogous quantity to the edge polarization will be the presence or absence of boundary SPT chains which we can indicate by the $\mathbb{Z}_2$ quantity $\chi^{edge}.$ Thus, we can define a higher-order topological indicator $\chi^{bulk}$ via (see Fig. \ref{fig:hotpdef}(b))
\begin{equation}
\chi^{bulk}=\chi^{corner}-\chi^{edge}_1-\chi^{edge}_2\,\mod\, 2
\end{equation} where $\chi^{edge}_{1,2}$ are the two topological indicators for the edges that intersect at the corner with indicator $\chi^{corner}.$ This definition implies that one has a non-trivial 2D HOTP in three different scenarios (i) both edges and the corner are topological, (ii) both edges are trivial and the corner is topological, or (iii) one edge is topological, and the other edge and the corner are trivial.

With this proposed definition it only remains to determine how one should explicitly calculate $\chi^{corner}$ and $\chi^{edge}$ given a model. In this article we can unambiguously calculate these quantities because our models are solvable and are in zero-correlation length limits so one can clearly identify the edge and corner degrees of freedom and make the proper assignment of the topological indicators. Indeed, we will find that our models fit scenario (i) above. For free fermion systems these quantities can be calculated even when away from such simple limits\cite{quadrupole,quadrupole2}, but calculating them in generic many-body systems remains an open problem. We will not solve this problem in this article, however we do provide numerical evidence in Appendix \ref{app:numerics} that our definition of a HOTP can be applied away from the zero-correlation length limit by analyzing the entanglement spectrum of the system to determine $\chi^{edge}$ similar to Ref. \onlinecite{ptbo}.

\section{$\mathbb{Z}_2\times\mathbb{Z}_2$ Cluster Model Construction}
\label{z2z2}
\emph{Review of 1d cluster model.} In Ref. \onlinecite{quadrupole}, the authors argued that, to generate a QTI, it was natural to start from a pair of dipoles in each unit cell such that the coupling of dipoles between unit cells would spread the dipoles to form a quadrupole. The dipole building block in the fermionic QTI was the Su-Schrieffer-Heeger insulating chain which is known to have a quantized dipole moment in the presence of certain symmetries. We take a similar approach here, where the fundamental building block for our 2d bosonic HOTP model is a 1d cluster model with two $\mathcal{Z}_2$ degrees of freedom per unit cell described by the following bosonic Hamiltonian:
\begin{equation}
  H_{\mathbb{Z}_2\times\mathbb{Z}_2}=-\sum_{i=1}^{N-1}\left(Z^a_{i}X^b_iZ^a_{i+1}+Z^b_iX^a_{i+1}Z^b_{i+1}\right)
  \label{z2z2_ham}
\end{equation}
where the $X^{a(b)}_i$ denotes a Pauli operator at a site $a(b)$ of the $i$-th unit cell. This model has a long history\cite{Suzuki1971,Raussendorf2001,Kopp2005}, and recent work \cite{Son2011} has shown this model to be in the non-trivial SPT phase protected by a global $\mathbb{Z}_2\times\mathbb{Z}_2$ symmetry.


  To see this, let us review some of the properties of this model. Consider a 1-dimensional chain with open boundary conditions. 
The Hamiltonian \ref{z2z2_ham} is a sum of $N-2$ commuting stabilizers $Z^a_{i}X^b_iZ^a_{i+1}$ and $Z^b_{i}X^a_{i+1}Z^b_{i+1}$ each of which squares to $1$. 
The ground state of such a system is a simultaneous eigenstate of all stabilizers with the eigenvalue $+1$. By counting the number of degrees of freedom and the number of stabilizers, we find that the ground state of this system (with open boundaries) must be 4-fold degenerate. 

To pin down the origin of this degeneracy, consider the following pair of symmetry operators:
\begin{equation}
  P_1=X^a_1X^a_2\dots X^a_{N},\quad P_2=X^b_1X^b_2\dots X^b_N.
  \label{eqn:z2z2_symmetries}
\end{equation}
The Hamiltonian commutes with, i.e., is symmetric under, the $\mathbb{Z}_2\times\mathbb{Z}_2$ group generated by $P_1$ and $P_2$. We can use the fact that, in the ground state, the commuting stabilizers all take the value $+1$ to  decompose both of these operators as a product of stabilizers which can be simplified to:
\begin{equation}
  \begin{split}
  &P_1=X^a_1Z^b_1\times Z^b_N\equiv P^L_1\times P^R_1\\ 
  &P_2=Z^a_1\times Z^a_{N}X^b_N\equiv P^L_2\times P^R_2.
\end{split}
  \label{eqn:factorized_symmetry2}
\end{equation}
Thus, despite the symmetry operators being global, they exhibit symmetry fractionalization to a pair of operators acting at each end of the chain in the ground state. Since the operators $P_1^L$ and $P_2^L$ that act on the same end anti-commute, we must have at least two degenerate modes at the left end, and similarly two degenerate modes at the right end, which means that four-fold degeneracy of the ground state comes from modes localized on the ends of the chain. 
The anti-commutation relations $P_1^{L(R)}P_2^{L(R)}=-P_2^{L(R)}P_1^{L(R)}$ are robust against any perturbation that respects the global $\mathbb{Z}_2\times\mathbb{Z}_2$ symmetry, though the forms of $P_{i}^{L(R)}$ may be modified. This implies that the end modes are protected by this symmetry, and the system is in nontrivial SPT phase.

\begin{figure}[h]
  \centering
  \includegraphics[width=0.5\textwidth]{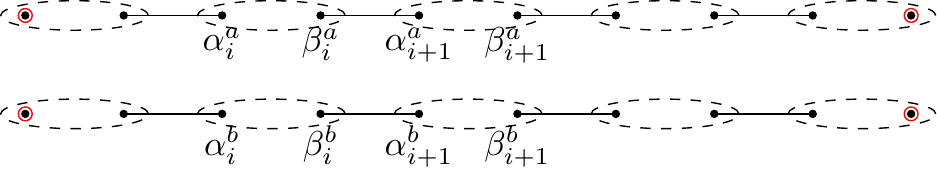}
  \caption{$\mathbb{Z}_2\times\mathbb{Z}_2$ spin chain Jordan-Wigner transformed to a pair of Kitaev chains in a topological phase. Each site of the original spin chain is represented by a pair of Majorana operators $\alpha_i$ and $\beta_i$ enclosed by an oval.}
  \label{fig:pair_of_chains}
\end{figure}

\emph{Majorana representation.} Since it will be useful for us later, let us review the representation of this model in terms of Majorana fermions using the following Jordan-Wigner transformation \cite{Fendley2012,iadecola2015} from one spin-$1/2$ degree of freedom to a pair of Majorana fermion operators:
\begin{equation}
\begin{split}
	&X^a_i=-i\alpha^a_i\beta^a_i,\quad X^b_i=-i\alpha^b_i\beta^b_i\\ 
    &Z^a_i=-i\prod_{j<i}(-\alpha^a_j\beta^a_j\alpha^b_j\beta^b_j)\alpha^a_i\\
    &Z^a_i=-i\prod_{j<i}(-\alpha^a_j\beta^a_j\alpha^b_j\beta^b_j)(-i\alpha^a_i\beta^a_i)\alpha^b_i.
    \label{eqn:jw_boson_to_fermion}
\end{split}
\end{equation}
This transformation transforms each stabilizer in our Hamiltonian into a  Majorana hopping term, and the total Hamiltonian \ref{z2z2_ham} takes the following form after the transformation:
\begin{equation}
  H_{\mathbb{Z}_2\times\mathbb{Z}_2}=-i\sum_{i=1}^{N-1}(\beta^a_{i}\alpha^a_{i+1}+\beta^b_{i}\alpha^b_{i+1}).
  \label{z2z2_Majorana}
\end{equation}
As we show in Fig. \ref{fig:pair_of_chains}, this Hamiltonian represents a pair of decoupled Kitaev chains \cite{Kitaev2001}. Furthermore, the four Majorana operators $\alpha^a_1$, $\alpha^b_1$, $\beta^a_{N}$, and $\beta^b_N$ are completely free, and reflect the end-mode degeneracy of the original model.
In the language of Majorana fermions, the edge modes are protected by time-reversal $T$ and total fermion parity $\mathcal{F}=\prod_{i=1}^{N}\alpha^a_{i}\beta^a_{i}\alpha^b_{i}\beta^b_{i}$ symmetries where: $T\alpha^{a(b)}_iT=\alpha^{a(b)}_i$ and $T\beta^{a(b)}_iT=-\beta^{a(b)}_i$ which is chosen such that it fixes the complex fermions $c^{a(b)}_i=\alpha^{a(b)}_i+i\beta^{a(b)}_i$ at each site to be invariant under $T=K$.  We could introduce couplings $i\alpha^a_1\alpha^b_1$ and $i\beta^a_{N}\beta^b_n$ to gap out the ends, but these terms break $T$; in the bosonic variables these terms translate to $Y^a_1Z^b_1=P_1^LP_2^L$ and $Z^a_{N}Y^b_N=P_1^RP_2^R$ which break the global $\mathbb{Z}_2\times\mathbb{Z}_2$ symmetry.

{\emph{2d HOTP Cluster Model.}} To construct our model we are going to couple a stack of \emph{pairs} of $\mathbb{Z}_2\times\mathbb{Z}_2$ cluster-model SPT chains. Explicitly, let us take a stack of $M$ pairs of $\mathbb{Z}_2\times\mathbb{Z}_2$ SPT chains of length $N$ oriented horizontally. We will only couple the pairs of chains \emph{between} unit cells in analogy with the zero correlation-length limit of the QTI model\cite{quadrupole}. Naturally, with this type of construction we will generate dangling $\mathbb{Z}_2\times\mathbb{Z}_2$ cluster-model chains on the top and bottom edges. Since the SPT classification of SPTs with this symmetry group is $\mathbb{Z}_2,$ we expect that after coupling pairs of these chains there will not be any gapless edge states, but the dangling chains on the edges could give rise to protected corner states. 

\begin{figure}
  \centering
  \includegraphics[width=0.5\textwidth]{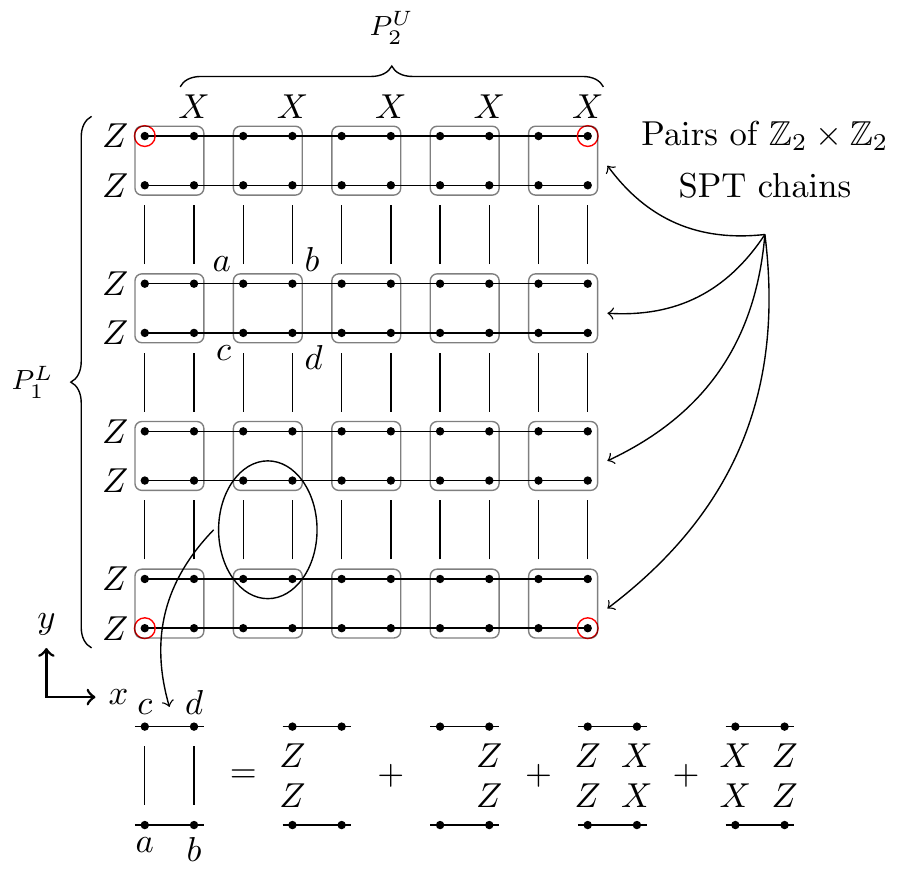}\\
  \caption{2-dimensional lattice built of pairs of $\mathbb{Z}_2\times\mathbb{Z}_2$ SPT chains. Vertical couplings are illustrated schematically below the lattice. This arrangement clearly leaves two decoupled SPT chains at the top and the bottom.}
  \label{fig:2dlattice}
\end{figure}

Our goal now is to choose the coupling in the vertical direction (between pairs of chains in neighboring unit cells) in such a way that the edge modes will be gapped, there will be non-trivial SPT chains on both the vertical and horizontal edges, and protected degeneracies on the corners.  We argue that the required coupling terms have the following form (see Fig. \ref{fig:2dlattice})
\begin{equation}
  \begin{split}
  &H_{V}=-\sum^{N}_{i=1} (Z^a_{i,A}Z^a_{i,B}+Z^b_{i,A}Z^b_{i,B}\\
  &+Z^a_{i,A}X^b_{i,A}Z^a_{i,B}X^b_{i,B}+X^a_{i,A}Z^b_{i,A}X^a_{i,B}Z^b_{i,B})
  \label{eqn:vertical_terms}
\end{split}
\end{equation}
where the pair of indices $i,j$ denotes the $i$-th unit cell of the $j$-th chain.

Since the terms only couple two chains, say $A$ and $B$, let us focus on just that single pair of chains for a moment. Introducing these terms breaks down the global symmetry group $\left(\mathbb{Z}_2\times\mathbb{Z}_2\right)^2\to\mathbb{Z}_2\times\mathbb{Z}_2$ with the new symmetry group generators taking the following form:
\begin{equation}
  P^{AB}_1=\prod_{i=1}^{N}X^a_{i,A}X^a_{i,B},\quad P^{AB}_2=\prod_{i=1}^{N}X^b_{i,A}X^b_{i,B}.  
  \label{eqn:symmetry_of_a_pair}
\end{equation}
The Hamiltonian for the pair of coupled chains is no longer a sum of the operators that all commute with each other, and the set of the conserved quantities are now represented by the products of two decoupled chains' stabilizers that became coupled to each other:
\begin{equation}
  \mathcal{O}^{AB}_i=Z^a_{i,A}X^b_{i,A}Z^a_{i+1,A}Z^a_{i,B}X^b_{i,B}Z^a_{i+1,B},
  \label{eqn:conserved_quantity}
\end{equation}
and a set of conserved quantities at the boundaries:
\begin{equation}
  \begin{split}
  &\mathcal{O}^{L,AB}_1=Z^a_{1,A}Z^a_{1,B},\quad \mathcal{O}^{L,AB}_2=X^a_{1,A}Z^b_{1,A}X^a_{1,B}Z^b_{1,B}\\
  &\mathcal{O}^{R,AB}_1=Z^b_{N,A}Z^b_{N,B},\quad \mathcal{O}^{R,AB}_2=Z^a_{N,A}X^b_{N,A}Z^a_{N,B}X^{b}_{N,B}.
  \label{eqn:boundary_conserved_quantities}
  \end{split}
\end{equation}
As was the case for a single chain, we can use the bulk conserved quantities to reduce the global symmetry to the product of boundary operators:
\begin{align}
 \begin{split}
  	P^{AB}_1&=X^a_{1,A} Z^b_{1,A} X^a_{1,B} Z^b_{1,B} \times Z^b_{N,A}Z^b_{N,B}\\
  	&=P^{L,AB}_1\times P^{R,AB}_1
	\end{split}
\\
\begin{split}
  P^{AB}_2&=Z^a_{1,A}Z^a_{1,B}\times Z^a_{N,A}X^b_{N,A}Z^a_{N,B}X^b_{N,B}\\
  &=P^{L,AB}_2\times P^{R,AB}_2.
\end{split}
  \label{eqn:2chain_symmetry_fractionalization}
\end{align}
However, here there is a key difference as the different operators acting on the same end of the doubled chain \emph{commute} with each other, signifying that the coupled pair of chains is in a trivial SPT phase with no required extra degeneracy from gapless modes present at the edges. Thus, we expect the edges of our 2d model to be gapped.

Now let us return to the full 2d model. Each unit cell on the lattice now consists of four $\mathbb{Z}_2$ degrees of freedom $a$, $b$, $c$, and $d$ as shown in Fig. \ref{fig:2dlattice}. After we collect the terms together, the full Hamiltonian reads: 
\begin{equation}
\label{eqn:H2d_total}
\begin{split}
	H_{2d}=&-\sum_{j=1}^{M}\sum_{i=1}^{N-1}\left(Z^a_{i,j}X^b_{i,j}Z^a_{i+1,j}+Z^b_{i,j}X^a_{i+1,j}Z^b_{i+1,j}\right.\\
    &\left.+Z^c_{i,j}X^d_{i,j}Z^c_{i+1,j}+Z^d_{i,j}X^c_{i+1,j}Z^d_{i+1,j}\right)\\
    &-\sum_{j=1}^{M-1}\sum_{i=1}^{N}\left(Z^a_{i,j}Z^c_{i,j+1}+Z^b_{i,j}Z^d_{i,j+1}\right.\\
  &\left.+Z^a_{i,j}X^b_{i,j}Z^c_{i,j+1}X^d_{i,j+1}+X^a_{i,j}Z^b_{i,j}X^c_{i,j+1}Z^d_{i,j+1}\right),
\end{split}
\end{equation}
and has a global $\mathbb{Z}_2\times\mathbb{Z}_2$ symmetry group generated by 
\begin{equation}
  P_1=\prod_{j=1}^M\prod_{i=1}^{N}X^a_{i,j}X^c_{i,j},\quad P_2=\prod_{j=1}^M\prod_{i=1}^{N}X^b_{i,j}X^d_{i,j}.
  \label{eqn:bulk_symmetries}
\end{equation} We want to show that (i) both horizontal and vertical edges are $\mathbb{Z}_2\times\mathbb{Z}_2$ SPT chains, and (ii) that there are protected corner modes that are shared by the two topological edges.  After we introduce vertical coupling terms, we automatically obtain that the horizontal 1-dimensional boundaries at the top and bottom are in a non-trivial SPT phase by the virtue of them being decoupled $\mathbb{Z}_2\times\mathbb{Z}_2$ SPT chains (see Fig. \ref{fig:2dlattice}). The symmetry generators for these two dangling chains (up and down) are inherited from the bulk global symmetry generators:
\begin{equation}
  \begin{split}
  &P^{up}_1=\prod_{i=1}^{N}X^a_{i,M},\quad P^{up}_2=\prod_{i=1}^{N}X^b_{i,M}\\
  &P^{down}_1=\prod_{i=1}^{N}X^c_{i,1},\quad P^{down}_2=\prod_{i=1}^{N}X^d_{i,1},\\
  \label{eqn:top_bottom_symmetry}
\end{split}
\end{equation}
where the index $(i,j)$ runs over lattice sites as indicated in Fig. \ref{fig:2dlattice}. As was discussed in the previous section, these symmetry operators fractionalize in the following way:
\begin{equation}
  \begin{split}
    &P^{up}_1=X^a_{1,M}Z^b_{1,M}\times Z^b_{N,M}\\ 
    &P^{up}_2=Z^a_{1,M}\times Z^a_{N,M}X^b_{N,M}\\
    &P^{down}_1=X^c_{1,1}Z^d_{1,1}\times Z^d_{N,1}\\ 
    &P^{down}_2=Z^c_{1,1}\times Z^c_{N,1}X^d_{N,1}.\\
  \label{eqn:top_bottom_fractionalized}
\end{split}
\end{equation} This gives four independent sets of anti-commuting, corner-localized algebras and thus at least a 16-fold ground state degeneracy. While this shows that there are protected corner modes, their presence is not equivalent to higher-order topology. We need to show that both vertical and horizontal pairs of edges are simultaneously topological.

Having seen that the horizontal edges are topological, we are now going to show that the vertical boundaries are also in a non-trivial SPT phase with gapless end modes that are shared with the horizontal boundary spin chains. Let us focus on the right boundary chain to be explicit. The Hamiltonian of this chain reads:
\begin{equation}
\label{HV_right}
H_{\mathbb{Z}_2\times\mathbb{Z}_2,R}=-\sum_{i=1}^{M-1}\left(Z^b_{i}Z^d_{i+1}+Z^a_{i}X^b_{i}Z^c_{i+1}X^d_{i+1}\right)
\end{equation}
where we drop the $\hat{x}$-coordinate index as it is fixed to be $N$ at the right boundary.
One can check that each individual term in this expression commutes not only with each other, but with the Hamiltonian of the total system (\ref{eqn:H2d_total}) as well. This allows us to immediately conclude that the chain on the right vertical edge is effectively decoupled from the rest of the system. This Hamiltonian looks quite different from Eq. \ref{z2z2_ham}, indeed the vertical and horizontal couplings of our model appear quite anisotropic at first sight. However, we can perform a unitary transformation that will map Eq. \ref{z2z2_ham} to Eq. \ref{HV_right}. This transformation is an effective $C_4$ rotation of the 2d HOTP model which performs the following transformation of the on-site operators: 
\begin{equation}
\begin{aligned}
\begin{split}
&Z^a\to\tilde{X}^c\tilde{Z}^d\\ 
&Z^b\to\tilde{Z}^a\\
&Z^c\to\tilde{Z}^d\\ 
&Z^d\to \tilde{Z}^a\tilde{X}^b
\end{split}
\end{aligned}
\qquad
\begin{aligned}
\begin{split}
&X^a\to\tilde{Z}^a\tilde{Z}^c\\
&X^b\to\tilde{X}^a\tilde{Z}^b\tilde{X}^c\tilde{Z}^d\\
&X^c\to\tilde{Z}^a\tilde{X}^b\tilde{Z}^c\tilde{X}^d\\
&X^d\to\tilde{Z}^d\tilde{Z}^b.
\end{split}
\end{aligned}
\end{equation}
This transformation maps the vertical edge Hamiltonian (\ref{HV_right}) to the familiar $\mathbb{Z}_2\times\mathbb{Z}_2$ cluster model Hamiltonian after we change the ordering direction:
\begin{equation}
\tilde{H}_{\mathbb{Z}_2\times\mathbb{Z}_2,R}=-\sum_{i=1}^{M-1}\left(\tilde{Z}^a_{i}\tilde{X}^b_{i}\tilde{Z}^a_{i+1}+\tilde{Z}^b_{i}\tilde{X}^a_{i+1}\tilde{Z}^b_{i+1}\right).
\end{equation}
Hence, we can conclude that vertical boundary chains are in the same SPT phase as the horizontal boundary chains.

In the original bosonic language, the symmetry operators for the left boundary chain of the lattice are:
\begin{equation}
  \begin{split}
  &P^{L}_1=\prod_{i=1}^{M}Z^a_{1,i}Z^c_{1,i}\\ 
  &P^{L}_2=\prod_{i=1}^{M}X^a_{1,i}Z^b_{1,i}X^c_{1,i}Z^d_{1,i}
  \label{eqn:left_vertical_symmetry}
\end{split}
\end{equation}
and for the right boundary chain:
\begin{equation}
  \begin{split}
  &P^{R}_1=\prod_{i=1}^{M}Z^b_{N,i}Z^d_{N,i}\\ 
  &P^{R}_2=\prod_{i=1}^{M}Z^a_{N,i}X^b_{N,i}Z^c_{N,i}
X^d_{N,i}  .
\label{eqn:right_vertical_symmetry}
\end{split}
\end{equation}
The set of edge conserved quantities for the paired chains (\ref{eqn:boundary_conserved_quantities}) allows us to fractionalize these symmetry operators as follows:
\begin{equation}
  \begin{split}
    &P^{L}_1=Z^c_{1,1}\times Z^a_{1,M}\\ 
    &P^{L}_2=X^c_{1,1}Z^d_{1,1}\times X^a_{1,M}Z^b_{1,M}\\
    &P^{R}_1=Z^d_{N,1}\times Z^b_{N,M}\\
    &P^{R}_2=Z^c_{N,1}X^d_{N,1}\times Z^a_{N,M}X^b_{N,M}.
\end{split}
  \label{eqn:vertical_symmetry_fractionalization}
\end{equation}
These operators  are exactly the same as the symmetry fractionalized operators for the top and bottom chains (\ref{eqn:top_bottom_fractionalized}), and we can conclude that \emph{the corner modes of the vertical chains are shared with the the horizontal chains}. 


\begin{figure}
  \centering
  \includegraphics[width=0.5\textwidth]{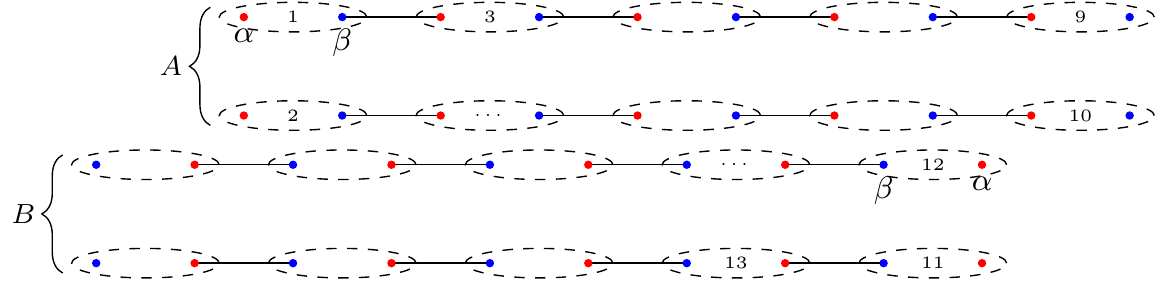}\\
  $\Downarrow$
  \includegraphics[width=0.5\textwidth]{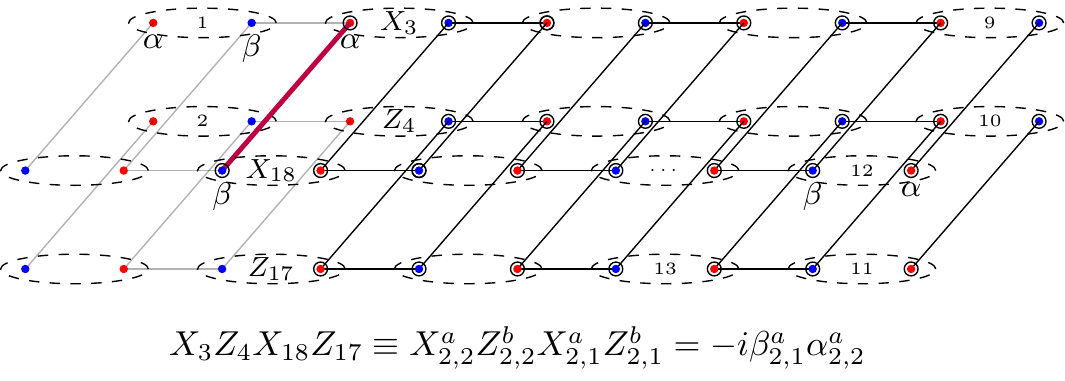}
  \caption{(upper) Two neighboring $\mathbb{Z}_2\times\mathbb{Z}_2$ spin chains (after Jordan-Wigner transformation to Majorana representation) that we intend to couple. (lower) Vertical coupling between two spin chains. In this particular example we focus on the coupling introduced by the $X_3Z_4X_{18}Z_{17}$ term, highlighted in purple. Dots enclosed by small circles represent the Majoranas engaged by this coupling term. Most of the Majoranas that enter the product can be removed using conserved quantities to leave simple quadratic hopping terms. Two dots connected by a purple bond represent the Majoranas that are left and cannot be combined into a conserved quantity.}
  \label{fig:four_chains}
\end{figure}

\emph{Majorana representation of the 2d lattice}
Each horizontal $\mathbb{Z}_2\times\mathbb{Z}_2$ chain can be Jordan-Wigner transformed as described in the beginning of this section where we reviewed the 1d cluster model. Since we are doing this on a 2d system there is a subtlety in the ordering when passing between the chains. We are going to choose the ordering to go in the opposite directions for two chains from the same pair: the upper chain ordering goes to the left and the lower chain ordering goes to the right, so that for a pair of chains that are coupled together, the ordering goes as depicted in the top of Fig. \ref{fig:four_chains}.


With this ordering, the Jordan-Wigner transformation converts the vertical coupling terms in the bosonic Hamiltonian into local Majorana hopping terms. Each of the eight terms in (\ref{eqn:H2d_total}) corresponds to a single hopping term after the transformation. As a result, the bulk of our model becomes a collection of clusters of four Majorana fermions with hopping between them as shown in Fig. \ref{fig:four_chains}.To understand how the coupling works, take a look at Fig. \ref{fig:four_chains}. Consider the operator $X_3Z_4X_{18}Z_{17}$ which can be written in terms of all of the encircled Majoranas in the figure:
\begin{equation}
X_3Z_4X_{18}Z_{17}=\alpha_3\beta_3\alpha_4\prod_{4<j<17}(-i\alpha_j\beta_j)\alpha_{17}\alpha_{18}\beta_{18}.
  \label{eqn:Z4_Z17}
\end{equation}
The product of four Majoranas that are connected by the black lines into a single plaquette in Fig. \ref{fig:four_chains} is a conserved quantity (\ref{eqn:conserved_quantity}). For example we have:
\begin{equation}
  \begin{split}
  Z_{5}X_{6}Z_{7}
Z_{14}X_{15}Z_{16}=-\beta_{5}\alpha_{7}\beta_{14}\alpha_{16}
\end{split}
\end{equation}
meaning that the product of Majoranas at the corners of one plaquette is a c-number, and we can conclude that $X_3Z_4X_{18}Z_{17}\propto i\alpha_{3}\beta_{18}$. The coefficient of proportionality depends on the value of the conserved quantities in a particular state. 

After carrying this out for each coupling term, we find that total model is a bilayer of Majorana fermions coupled on plaquettes with hopping terms. To corroborate that our model represents a HOTP we want to confirm that each plaquette contains an effective $\pi$-flux\cite{quadrupole}. The $\pi$-flux will be represented by the relative signs of the various hopping terms, and these are determined by the conserved quantities in the ground state. We checked the values of the conserved quantities in the ground state  and find that the four hopping terms for a single plaquette after the Jordan-Wigner transformation are:
\begin{equation}
  \begin{split}
  H_{p}=&-i\beta^c_{i,j+1}\alpha^c_{i+1,j+1} + i\alpha^a_{i,j}\beta^a_{i+1,j} \\
  &-i\alpha^a_{i,j}\beta^c_{i,j+1}-i\beta^a_{i+1,j}\alpha^c_{i+1,j+1}.
\end{split}
  \label{eqn:h_quad}
\end{equation}
This leads us to conclude that the effective flux through each plaquette is $\pi$. Thus, the bosonic system can be represented in terms of two copies of the higher order topological superconductor discussed in Ref. \onlinecite{Wang2018}. The edges of the model are a pair of gapped Kitaev chains, and there is a (symmetry-protected) pair of free Majorana modes on each corner of the system that gives rise to the two-state Hilbert space of the dangling spin in the bosonic degrees of freedom.

\begin{figure}
	\centering
        \includegraphics[width=0.4\textwidth]{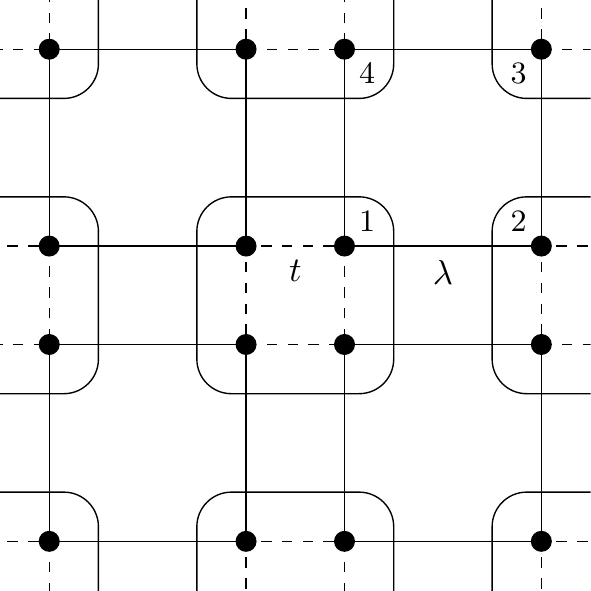}
	\caption{Square lattice with four spin-$1/2$ degrees of freedom per unit cell coupled via dimerized antiferromagnetic XY couplings.} 
	\label{quadrupole}
\end{figure}

\section{XY model}\label{xysec}
We now want to provide a second model, with more conventional spin-spin couplings, that will exhibit the same bosonic HOTP. The $\mathbb{Z}_2\times\mathbb{Z}_2$ cluster model can be deformed to a more conventional model of a dimerized spin-$1/2$ Heisenberg or XY chain\cite{Verresen2017}. Thus, we expect that instead of taking the cluster-model couplings we can instead create a very simple model of a dimerized XY spin system. Note that the discussion below can be generalized to Heisenberg couplings with minor modifications to the equations below. 

Explicitly, let us consider a square lattice with four spin-$1/2$ degrees of freedom per unit cell. The lattice structure is the same as in the QTI\cite{quadrupole}, and is depicted in Fig. \ref{quadrupole}. We couple neighboring spins via antiferromagnetic XY interactions with alternating coupling constants $t$ and $\lambda$ (both greater than zero) for the intra- and inter-cell interactions respectively. To begin with, consider the XY model on a finite $N\times N$ lattice and set the value of intra-cell coupling $t$ to zero. In this limit, the bulk of the lattice is a collection of decoupled clusters of four spins connected via XY interactions, and the edges are reduced to a collection of dimers. The horizontal and vertical edges are gapped, and are dimerized, antiferromagnetic spin-$1/2$ chains in a non-trivial $\mathbb{Z}_2\times\mathbb{Z}_2$ SPT state (the details of the symmetry will be discussed below). Furthermore, we also find four spins at the corners of the lattice that are completely free in this limit. Hence, we see all of the phenomenology expected for a HOTP.

From the structure of the couplings we see that the ground state in the $t=0$ limit is quite simple to compute as it is just a product state of the $4$-spin cluster ground states on each bulk site, the dimer singlet ground state for each edge site, and the degenerate corner spins. For the $4$-spin clusters, the Hamiltonian for each cluster $p$ is:
\begin{equation}
 H_p=\lambda\sum_{a=x,y}\left(\sigma^a_1\sigma^a_2+\sigma^a_2\sigma^a_3+\sigma^a_3\sigma^a_4+\sigma^a_4\sigma^a_1\right)
\end{equation}
and the ground state can be computed to be:
\begin{equation}
\begin{split}
|GS\rangle_p=&\frac{1}{2\sqrt{2}}\left(|\uparrow\uparrow\downarrow\downarrow\rangle+|\downarrow\uparrow\uparrow\downarrow\rangle+|\downarrow\downarrow\uparrow\uparrow\rangle+\right.\\
&\left.|\uparrow\downarrow\downarrow\uparrow\rangle\right)-\frac{1}{2}\left(|\downarrow\uparrow\downarrow\uparrow\rangle+|\uparrow\downarrow\uparrow\downarrow\rangle\right).
\end{split}
\end{equation} The energy gap to the first excited cluster state is $\Delta_p=4(\sqrt{2}-1)\lambda.$
Each of the boundary dimers $d$ is described by a simple Hamiltonian with a well-known ground state:
\begin{align}
  &H_{d}=\lambda \sum_{a=x,y}\sigma^a_1\sigma^a_2\\
&|GS\rangle_d=\frac{1}{\sqrt{2}}\left(|\uparrow\downarrow\rangle-|\downarrow\uparrow\rangle\right).
\end{align} The energy gap to the excited dimer state is $\Delta_d=2\lambda.$

Thus, in the absence of any corners, e.g., with fully periodic boundary conditions, the system is gapped and has a unique ground state. In the presence of four corners  we deduce that the ground state degeneracy of the whole system is $2^4=16$. If we tune away from the $t=0$ limit the degeneracy is still preserved, and any splitting is exponentially small in the system size. To see this explicitly we performed degenerate perturbation theory on the ground state subspace and found that the effective coupling between corner spins scales as $(t/\lambda)^{N/2}$ and thus is exponentially small as $N$ increases (see Appendix \ref{perturb}).

The global $\mathbb{Z}_2\times\mathbb{Z}_2$ symmetry group can be generated by $\pi$ spin rotations around the $x$ and $y$ axes. To be explicit consider the global $\mathbb{Z}_2\times\mathbb{Z}_2$ symmetry group of the XY model generated by:
\begin{equation}
\label{XY_symmetry}
P_1=\prod_{\textbf{i}}\sigma^x_{\textbf{i}},\quad P_2=\prod_{\textbf{i}}\sigma^y_{\textbf{i}}
\end{equation}
where the index $\textbf{i}$ runs over each spin of the lattice. To see the symmetry fractionalization, consider the XY model in a zero correlation length limit ($t=0$). In this limit the model admits a set of conserved quantities associated with every bulk plaquette $p$ and edge dimer $d$:
\begin{equation}
\begin{split}
\mathcal{O}^1_p=\prod_{\textbf{i}\in p}\sigma^x_{\textbf{i}},\quad \mathcal{O}^2_p=\prod_{\textbf{i}\in p}\sigma^y_{\textbf{i}}\\
\mathcal{O}^1_d=\prod_{\textbf{i}\in d}\sigma^x_{\textbf{i}},\quad \mathcal{O}^2_d=\prod_{\textbf{i}\in d}\sigma^y_{\textbf{i}}
\end{split}
\end{equation}
Much like in the case of 2d HOTP Cluster Model, the set of conserved quantities allows us to reduce the symmetry group generators (\ref{XY_symmetry}) to the product of four corner-localized operators on the $N\times N$ lattice:
\begin{equation}
\label{eqn:XY_symmetry_fractionalized}
\begin{split}
P_1=\sigma^x_{1,1}\times\sigma^x_{1,N}\times\sigma^x_{N,1}\times\sigma^x_{N,N}\\ 
P_2=\sigma^y_{1,1}\times\sigma^y_{1,N}\times\sigma^y_{N,1}\times\sigma^y_{N,N}
\end{split}
\end{equation}
Once again, we have obtained a set of four anti-commuting, corner-localized algebras that signify the presence of the degeneracy from the gapless corner modes.

\emph{Jordan-Wigner transformation of the XY model.}
To further elucidate the connection of the XY model to the fermion quadrupole model we are going to use a Jordan-Wigner transformation presented in Ref. \onlinecite{wang1991}. In this subsection we are working on a square lattice with one spin per site. We start, as usual, by identifying spin ladder operators with fermionic creation and annihilation operators with position-dependent phase factors:
\begin{equation}
\label{WangJW}
S^-_i=\text{e}^{i\phi_i}d_i,\quad S^+_i=\text{e}^{-i\phi_i}d^\dagger_i
\end{equation}
where the phase factors and ladder operators are defined as:
\begin{equation}
\phi_i=\sum_{j\neq i}B_{ij}d_j^\dagger d_j,\quad S^\pm=S^x\pm iS^y
\end{equation}
The condition for $d_i$ and $d_i^\dagger$ to obey the correct fermionic statistics can be expressed via $B_{nm}$. It reads:
\begin{equation}
\text{e}^{iB_{nm}}=-\text{e}^{iB_{mn}}.
\end{equation}
As long as this condition is satisfied, the Jordan-Wigner transformation functions correctly, however, there is one particularly useful choice of $B_{mn}$:
\begin{equation}
  B_{mn}=\arg(\vec{r}_m-\vec{r}_n)=\mbox{\rm Im}\ln(z_m-z_n)
\end{equation}
where we have defined a complex coordinate at each site $\textbf{j}=(x_j,y_j)$: $z_j=x_j+iy_j$. This choice of $B_{mn}$ leads to the following phases $\phi_i$:
\begin{equation}
  \label{phase}
  \phi_i=\sum_{j\neq i}\mbox{\rm Im}\ln(z_j-z_i) d_j^\dagger d_j.
\end{equation}
Now, consider the spin-1/2 XY-model Hamiltonian with alternating coupling constants:
\begin{equation}
  H_{XY}=\sum_{\langle \textbf{i}\textbf{j}\rangle}J_{\textbf{i},\textbf{j}}\sum_{a=x,y}\sigma^a_{\textbf{i}}\sigma^a_{\textbf{j}}
\end{equation}
where $J_{\textbf{i}\textbf{j}}$ along $\hat{x}$ equals $\lambda$ for even values of the coordinate $x$, and $t$ for odd, and similarly along $\hat{y}$. After applying the transformation (\ref{WangJW}) we obtain the following Hamiltonian:
\begin{equation}
  \begin{split}
    H=&\sum_{\langle \textbf{i}\textbf{j}\rangle}\frac{1}{2}J_{\textbf{i},\textbf{j}}\left(d^\dagger_\textbf{i}\text{e}^{-i(\phi_i-\phi_j)}d^{\phantom{\dagger}}_\textbf{j}+d^{\phantom{\dagger}}_\textbf{i}\text{e}^{i(\phi_i-\phi_j)}d^\dagger_\textbf{j}\right).  
    \end{split}
\end{equation}
This is a free-fermion Hamiltonian in an external gauge field, where
\begin{equation}
  \phi_i-\phi_j=\int_{\textbf{j}}^{\textbf{i}}\textbf{A}\cdot\text{d}\textbf{r}.
\end{equation}
We can substitute (\ref{phase}) and solve for $\textbf{A}$ to find:
\begin{equation}
\label{A_JW}
  \textbf{A}(\textbf{r})=-\sum_{\textbf{r}'\neq\textbf{r}}n_{\textbf{r}'}\frac{\hat{z}\times(\textbf{r}'-\textbf{r})}{(\textbf{r}'-\textbf{r})^2}
\end{equation}
where $n_{\textbf{r}}=d^\dagger_{\textbf{r}}d_{\bf{r}}=S^+_{\bf{r}}S^-_{\bf{r}}=\frac12+S^z$. In the absence of the external magnetic field we can make the following mean-field approximation:
\begin{equation}
n_{\textbf{r}}\to\langle n\rangle=\frac{1}{2}+\langle S^z\rangle=\frac12.
\end{equation}
We can justify this step in the zero correlation length limit, when $t=0$ and the bulk of the system is simply a collection of disjoint clusters of four spins. In this limit we can directly compute that $\langle S^z\rangle=0$ in the ground state.
By taking a continuum limit, we express the sum in \ref{A_JW} as:
\begin{equation}
\textbf{A}(\textbf{r})=-\frac{\langle n\rangle}{S}\int d^2\textbf{r}'\frac{\hat{z}\times(\textbf{r}'-\textbf{r})}{(\textbf{r}'-\textbf{r})^2}=\pi\frac{\langle n\rangle}{S}\ \hat{z}\times\textbf{r}
\end{equation}
where $S$ is the area of an elementary plaquette. This allows us to compute the magnetic field:
\begin{equation}
\textbf{H}(\textbf{r})=\text{rot}\textbf{A}(\textbf{r})=2\pi \frac{\langle n\rangle}{S}\hat{z}
\end{equation}
which leads to the flux per plaquette:
\begin{equation}
\Phi=\textbf{H}(\textbf{r})\cdot \hat{z}\cdot S=2\pi\langle n\rangle=\pi.
\end{equation}
This is exactly the condition that was imposed in the original fermionic QTI model\cite{quadrupole} for it to be in a topologically non-trivial phase. And the complete Hamiltonian which is Jordan-Wigner dual to spin Quadrupole model is:
\begin{equation}
  H_{XY}=H_{QP}
\end{equation}
where $H_{QP}$ is the fermionic QTI model Hamiltonian in the zero correlation length limit.

Applying this Jordan-Wigner transformation to the fractionalized $\mathbb{Z}_2\times\mathbb{Z}_2$ group generators (\ref{eqn:XY_symmetry_fractionalized}) and rewriting $d_i$ and $d_i^\dagger$ via Majorana operators we find:
\begin{equation}
\label{eqn:XY_sincos}
\begin{split}
&P_1=\prod_{\textbf{i}}\left(\cos(\phi_{\textbf{i}})\alpha_{\textbf{i}}-\sin(\phi_{\textbf{i}})\beta_{\textbf{i}}\right)\\
&P_2=\prod_{\textbf{i}}\left(-\sin(\phi_{\textbf{i}})\alpha_{\textbf{i}}-\cos(\phi_{\textbf{i}})\beta_{\textbf{i}}\right)
\end{split}
\end{equation}
where the index $\textbf{i}$ runs over four corner sites $(1,1)$, $(N,1)$, $(N,N)$, and $(1,N)$ which we denote below as $1$, $2$, $3$, and $4$ respectively. Majorana fermions commute with the phase factors in the following way: 
\begin{equation}
\begin{split}
&\cos(\phi_i)\gamma_i=\gamma_i\cos(\phi_i)\\
&\cos(\phi_i)\gamma_j=\gamma_j\cos(\phi_i-2B_{ij}n_j)
\end{split}
\end{equation}
and similarly for $\sin(\phi_i)$. 
With the help of these commutation relations we can move all the phase factors to the right side of every term in the product (\ref{eqn:XY_sincos}).
Now we can fix the gauge, requiring for the phases
\begin{equation}
\begin{split}
&\phi_1-2B_{12}n_2-2B_{13}n_3-2B_{14}n_4,\\
&\phi_2-2B_{23}n_3-2B_{24}n_4,\\
&\phi_3-2B_{34}n_4,\ \text{and}\  \phi_4
\end{split}
\end{equation}
to be the multiples of $2\pi$ thus drastically simplifying the form of the symmetry operators:
\begin{equation}
P_1=\alpha_1\alpha_2\alpha_3\alpha_4,\quad P_2=\beta_1\beta_2\beta_3\beta_4.
\end{equation}
Once again, we brought our symmetry operators to the form that makes the presence of gapless corner modes evident.
This form of $\mathbb{Z}_2\times\mathbb{Z}_2$ group generators also makes it clear that an inclusion of a magnetic field $h\sum_{i} \sigma^z_{i}\propto h\sum_{i}(\alpha_{i}\beta_{i}+...)$, which breaks the time-reversal symmetry, would break the global $\mathbb{Z}_2\times\mathbb{Z}_2$ symmetry generated by a pair of operators above. 

\emph{3d XY model}
We can construct a simple extension of the XY model to 3d.
Consider a cubic lattice with eight spin-1/2 degrees of freedom per unit cell. The lattice has alternating coupling strengths in all three directions repeating the structure of the 3d topological octupole insulator\cite{quadrupole}. 
We couple neighboring spins via antiferromagnetic XY interactions with coupling strength $t$ for spins at the same unit cell and coupling strength $\lambda$ for a pair of spins from different unit cells.
As we did for the 2d XY model, let us consider the 3d model on a finite $N\times N\times N$ lattice with the value of intra-cell couplings $t$ set to zero. 
In this limit, the bulk of the lattice decouples into a collection of clusters of eight spins connected via XY interactions while on each face of the lattice we find the already familiar 2d XY model. 
The boundary of the 3d XY model is gapped as each face of the lattice is in the 2d HOTP, and each of the twelve edges is in the non-trivial $\mathbb{Z}_2\times\mathbb{Z}_2$ SPT phase. 
Finally, we find eight spins at the corners of the lattice to be completely free in the $t=0$ limit. 

We can directly compute the ground state of a single bulk cluster to find it to be unique with the energy gap equal to $\Delta_c\approx 1.056\ \lambda$.
Thus, we can conclude that this model has a unique ground state on a lattice with periodic boundary conditions in all spatial directions. Considering the 3d XY model on a finite lattice, we find the ground state degeneracy to be $2^8=256$-fold. 
As in the 2d XY model, this degeneracy remains robust even as we tune away from the $t=0$ limit. Applying degenerate perturbation theory we find that the effective coupling between the corner spins decreases exponentially with the system size $t_{eff}\sim (t/\lambda)^N$ as detailed in Appendix \ref{perturb}.

The global $\mathbb{Z}_2\times\mathbb{Z}_2$ symmetry group can be generated by a pair of $\pi$ rotations of every spin on the lattice around the $x$ and $y$ axes:
\begin{equation}
\label{3dXY_symmetry}
P_1=\prod_{\textbf{i}}\sigma^x_{\textbf{i}},\quad P_2=\prod_{\textbf{i}}\sigma^y_{\textbf{i}}
\end{equation}
with the index $\textbf{i}$ running over each spin of the lattice. These symmetry operators can be fractionalized. In the $t=0$ limit, in addition to the set of boundary conserved quantities of the 2d XY model, we have a set of conserved quantities associated with each bulk cluster $c$ of eight spins:
\begin{equation}
\mathcal{O}^1_c=\prod_{\textbf{i}\in c}\sigma^x_{\textbf{i}},\quad \mathcal{O}^2_c=\prod_{\textbf{i}\in c}\sigma^y_{\textbf{i}}
\end{equation}
As in the 2d XY model, the set of conserved quantities allows us to reduce our pair of symmetry operators (\ref{3dXY_symmetry}) to the products of eight corner spin rotations:
\begin{equation}
\label{eqn:XY_symmetry_fractionalized}
\begin{split}
P_1=\sigma^x_{1,1,1}\sigma^x_{1,1,N}\sigma^x_{1,N,1}\sigma^x_{1,N,N} \sigma^x_{N,1,1}\sigma^x_{N,1,N}\sigma^x_{N,N,1}\sigma^x_{N,N,N}\\ 
P_2=\sigma^y_{1,1,1}\sigma^y_{1,1,N}\sigma^y_{1,N,1}\sigma^y_{1,N,N} \sigma^y_{N,1,1}\sigma^y_{N,1,N}\sigma^y_{N,N,1}\sigma^y_{N,N,N}
\end{split}
\end{equation}
Thus we obtain eight corner-localized, anti-commuting algebras that indicate the existence of gapless corner modes spawning the $2^8$-fold degeneracy.

\section{Conclusion}\label{conclusion}
We have presented two separate models that realize a $\mathbb{Z}_2\times\mathbb{Z}_2$ higher order topological phase. Despite their different microscopic Hamiltonians they yield similar phenomenology with gapped edges realizing 1d SPT phases and protected modes localized on the corners. The concepts presented here can be generalized to other symmetry groups (including spatial symmetries) and spatial dimensions. 

We provided an extension of the XY model to 3d constructing a third-order HOTP by forming XY coupled cubes of spins in the 3d bulk in analogy with the coupled plaquettes in the bulk of the 2d XY model. One missing construction in this article is a model with conventional spin interactions for a second-order phase in 3d (though Refs. \onlinecite{Song16,yizhiarxiv} for example, show interacting models with hinge states) and it would be interesting to see if one can create such a model by stacking and coupling our models for 2d second order phases. 

Perhaps the largest open question is how to generically calculate the edge and corner topological indicators that serve to define the topological phase in a many-body system. It is possible that the entanglement spectrum may provide a route to do this similar to the use of Wannier bands for free fermion systems. The resolution of this issue will provide a major advance in understanding many-body higher order topological phases. 


{\bf{Note:}} During the preparation of this manuscript we became aware of a partially overlapping and complementary article by Yizhi You, Trithep Devakul, Fiona J. Burnell, Titus Neupert\cite{yizhiarxiv}. 

\begin{acknowledgments}
We thank W. A. Benalcazar, B. A. Bernevig, F. J. Burnell, M. Hermele, L. Santos, and Y. You for useful discussions. TLH acknowledges support from the US National Science
Foundation under grant DMR 1351895-CAR.
\end{acknowledgments}

\bibliography{lib.bib}

\appendix

\section{Degenerate perturbation theory}\label{perturb}
\emph{Effective Hamiltonian}
Consider a Hamiltonian of the form:
\begin{equation}
	H=H_0+gH_1.
	\label{pert_ham}
\end{equation}
Suppose that $H_0$ has a degenerate ground state with the energy $E_0$. We want to write down an effective Hamiltonian that will describe energy splitting of the ground state due to the perturbation $H_1$. To do so, we will briefly recount the derivation. Introducing the set of projectors $P_n$ into the degenerate-energy $E_n$ subspaces of the Hilbert space $\mathcal{H},$ we can write the unperturbed Hamiltonian spectral decomposition:
\begin{equation}
	H_0=\sum_nE_nP_n.
	\label{spectral}
\end{equation}

We aim to find an effective Hamiltonian for the degenerate ground state, therefore, we should slightly modify the usual Schr\"odinger equation:
\begin{equation}
\begin{split}
	&H|\psi\rangle=E|\psi\rangle \ \to\ H_{eff}|\tilde{\psi}\rangle=E|\tilde\psi\rangle\ \\ 
	&\text{where}\ |\tilde\psi\rangle=P_0|\psi\rangle
\end{split}
\end{equation} 

Introducing the operator $P_\perp=1-P_0$ we can write down an effective Hamiltonian $H_{eff}$ in the following form:
\begin{equation}
	H_{eff}=P_0H\left(1+(E-P_\perp H)^{-1}P_\perp H\right)P_0
\end{equation}
substituting here (\ref{pert_ham}) and (\ref{spectral}) we obtain the following form of the effective Hamiltonian for Hilbert subspace generated by $|\tilde\psi\rangle$:
\begin{equation}
	H_{eff}=E_0+PH_1\sum_{n=0}^\infty g^{n+1}\left(\sum_{m>0}\frac{P_mH_1}{E-E_m}\right)^n
	\label{eff_ham}
\end{equation}
where $E=E_0+\sum_{n=1}^\infty g^n E^{(n)}$.
Simply put, to obtain the $n$-th order of $H_{eff}$ we apply perturbation operator $H_1$ to the ground state and project the resulting vector back to the ground state subspace. 

\emph{Perturbative stability of the corner modes.}
Now, let us show that the corner mode degeneracy is protected in the thermodynamic limit when we are away from the zero correlation-length limit, i.e., it only has exponentially small splitting in finite sized systems.
To do so, we will study the 3d dimerized XY model. The reasoning in this section can be easily applied to the 2d XY model. 
We start by turning the intra-cell XY interaction terms $t$ to be non-zero. Assuming that $\lambda\gg t$ we can employ the degenerate perturbation theory described above to obtain an effective Hamiltonian for the ground state, as well as the ground state energy splitting. 

The $n$-th order contribution to the effective Hamiltonian is given by:
\begin{equation}
  H^{(n)}=P_0V\left(\sum_{m>0}\frac{P_m H_1}{E-E_m}\right)^n
  \label{eqn:nth_order_ham}
\end{equation}
where $P_0$ is the projector to the ground state subspace, $P_m$ is the projector to the $m$-th energy subspace and $H_1$ is the perturbation, which, in our case, is given by the sum of all $t\sum_{a=x,y}\sigma^a_i\sigma^a_j$ terms where spins $i$ and $j$ belong to the same unit cell. To obtain an effective Hamiltonian for the corner modes we need to find a combination of perturbation terms that: (a) act non-trivially on the corner spins and (b) the resulting state has a non-zero projection back to the ground state subspace. While the first criteria is easy to satisfy simply by starting with the perturbation term from one of the corner unit cells, the second criteria requires that every perturbed (bulk) plaquette or (edge) dimer wavefunction has a non-zero projection to the ground state. Each intra-cell XY perturbation term acts on two separate spin clusters (either bulk cubes, face plaquettes, edge dimers or corner spins), and while the corner spins always have a non-zero projection back to the ground state, cubes, plaquettes, and dimers must be acted upon by at least two different perturbation terms with the same $\sigma^a$ operators to have a non-zero projection to the ground state. 

The latter requirement leads us to the conclusion that the effective Hamiltonian is represented by products of $\sigma^a_i$ that form either closed loops, where each perturbed spin cluster is being acted upon by two distinct $\sigma^a_i\sigma^a_j$ operators, or strings of $\sigma^a_i$ that stretch from one corner of the lattice to another. Any allowed loop operator acts trivially on the corner state so the effective Hamiltonian for the corner spins is given by  $t_{eff}\sigma^a_i\sigma^a_j$ where the first and the last $\sigma^a$ act on different corner spins. The number of perturbation operators involved is at least $N/2$ and the resulting effective Hamiltonian has a coupling constant of roughly $t_{eff}\sim(t/\lambda)^{N/2}$ rendering ground state degeneracy to be exponentially protected by the system size $N$.

\section{Numerical Calculations}\label{app:numerics}
Following Ref. \onlinecite{ptbo} we are going to numerically examine the entanglement structure of the XY model. The degeneracy of the low-lying entanglement spectrum was proposed as a fingerprint to identify topological order by Li \& Haldane in Ref. \onlinecite{LiHaldane}. Two-dimensional models are difficult to study via exact diagonalization, so we resort to the iTEBD technique \cite{itebd} to confirm the four-fold degeneracy of the ground state in the limit when $t$ is significantly smaller than $\lambda$. Qualitatively, our results hold for the Heisenberg model as well and we will use this model to illustrate certain points as some aspects of the numerical simulation are more pronounced in the Heisenberg case rather than in the XY model. 

\begin{figure}[h]
	\centering
        \includegraphics[width=0.45\textwidth]{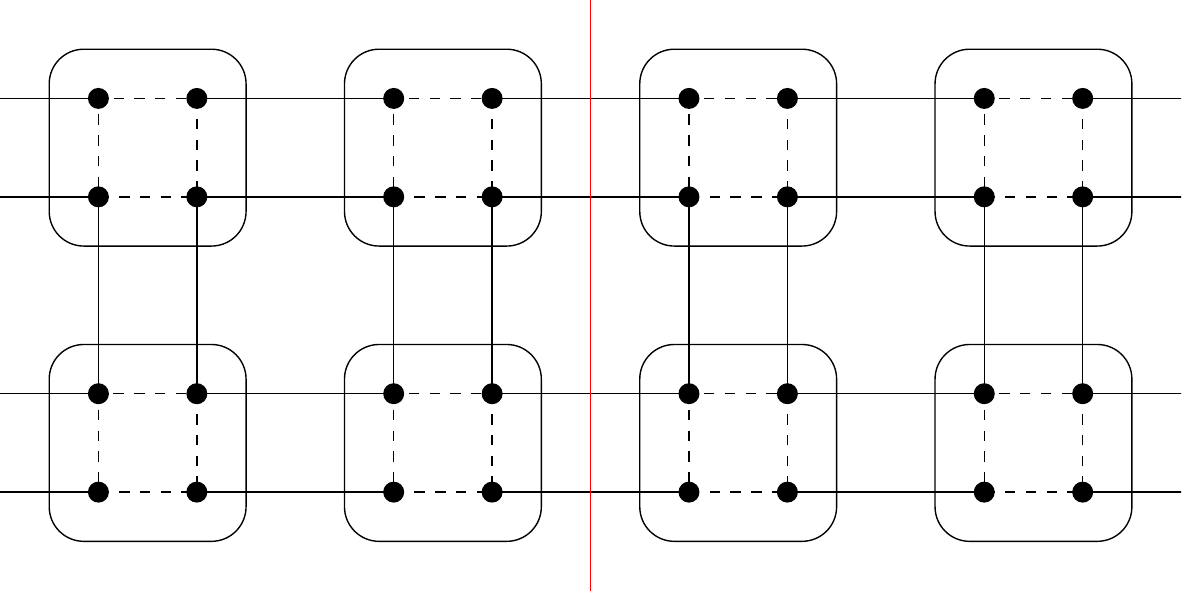}
	\caption{Quadrupole $L_y=2$ strip with the entanglement cut} 
	\label{qp_1Ditebd}
\end{figure}

\begin{figure}
\begin{tikzpicture}
    \node[anchor=south west,inner sep=0] at (0,0) {\includegraphics[width=0.45\textwidth]{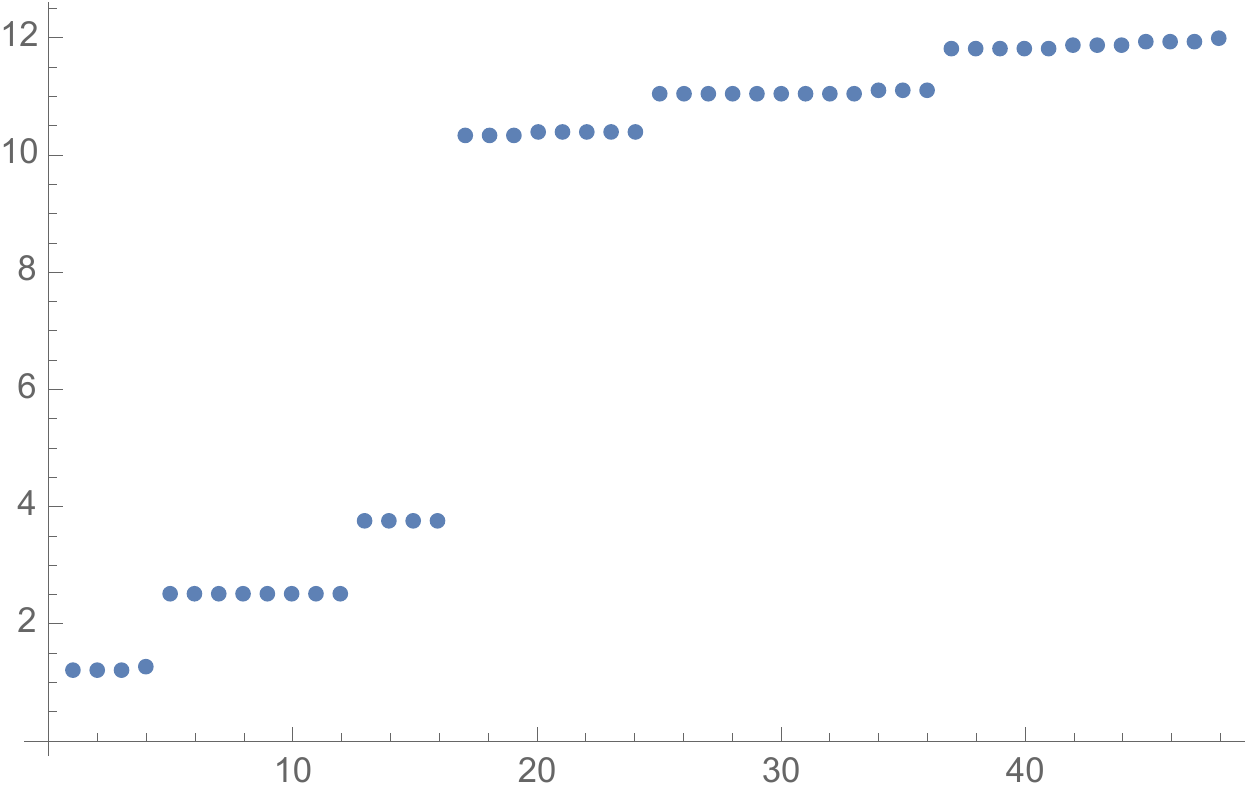}};
    \node [rotate=90] at (-0.4,3) {$-2\log\lambda_n$}; 
    
\end{tikzpicture}
\caption{Entanglement spectrum calculated for $t_x=t_y=0.1$}
\label{fig:entsp_28}
\end{figure}

\emph{Quasi-1d Calculations}
The iTEBD algorithm is very useful when it comes to calculation of the entanglement spectrum in 1d models. We are working with 2d model, however we are interested in computing the entanglement spectrum for a simple cut that splits our lattice in half, eliminating a set of horizontal $\lambda$ bonds as shown in Fig. \ref{qp_1Ditebd}. The goal is to confirm that the boundary SPT chains remain even away from the zero-correlation length limit. If we now consider a spin quadrupole model defined on an infinite strip of width $L_y=2$ we can apply the 1d version of iTEBD algorithm by treating four spins with the same $\hat{x}$ coordinate as being located on the same site. 
This setup allows us to simulate an infinite lattice along $\hat{x}$ direction meaning that the ground state degeneracy is only lifted via $t$ couplings along $\hat{y}$ direction. The entanglement spectrum for $t=0.1$ is presented in Fig. \ref{fig:entsp_28}.


Numerical data is presented in table below. The iTEBD was implemented with the bond dimension equal to 28. Inter-cell coupling was fixed to be $\lambda=1$ and $t$ varied between $0.01$ and 0.3. As can be seen from the table below, initially the entanglement spectrum is four-fold degenerate as was computed exactly for the analytic ground state wavefunction for the case with $t=0$. This degeneracy is coming from the two non-trivial SPT chains that we have cut on the top and the bottom; each of which yields two degenerate entanglement modes. In our numerics, the splitting of the degeneracy becomes more pronounced starting near $t=0.3,$ but the lowest modes are still well-separated from the first set of excited states. The entanglement spectrum for different values of $t$ is presented in Table \ref{entsp_table}.

\begin{figure}
	\centering
        \includegraphics[width=0.45\textwidth]{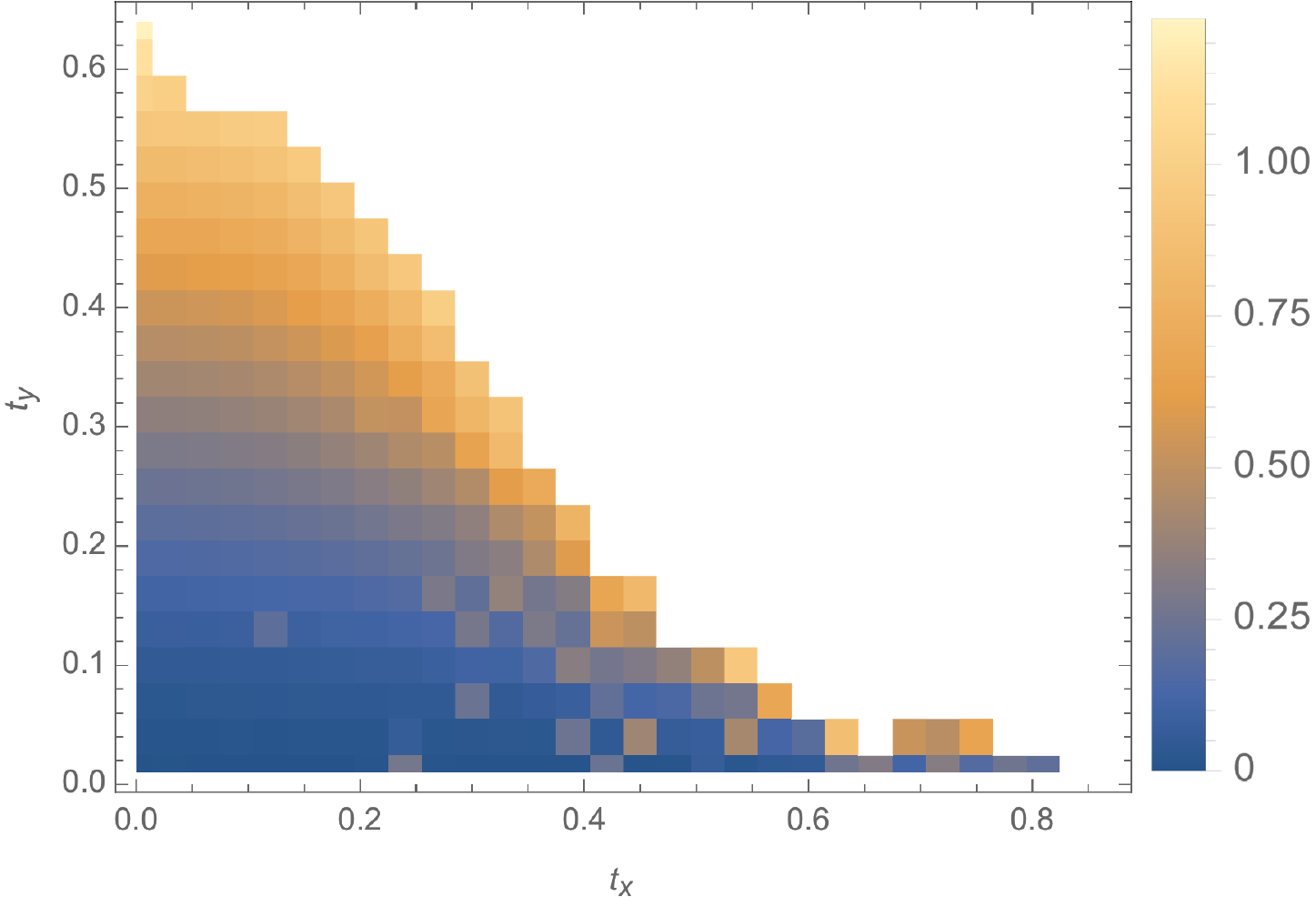}
	\caption{Degeneracy of the $L_y=4$ quadrupole strip for different values of $t_x$ and $t_y$. White area corresponds to the parameters' values for which we can't reliably apply the iTEBD algorithm without drastically increasing the bond dimension.} 
	\label{entsp_heatmap}
\end{figure}

There are several consistency checks for this model that we carried out. First of all, introducing periodic boundary conditions in the vertical direction immediately breaks the even degeneracy of the spectrum even for $t=0$. Additionally, we quantified the four-fold degeneracy for different values of intra-cell couplings  $t_x$ and $t_y$ in different directions. The results of this calculation can be seen on the Figure \ref{entsp_heatmap}. In this case the iTEBD was implemented with bond dimension $\chi=48$. Values of $t_x$ and $t_y$ run from 0.01 to 0.75 and 0.01 to 0.65 respectively. As a measure of degeneracy breakdown of the ground state we took the difference of the first and fourth eigenvalue (bandwidth of ground state splitting) relative to the difference of the fifth and fourth eigenvalue (gap to excited states). As evident from the diagram, for the small values of $t_x$ and $t_y$ the degeneracy is almost unbroken but if we start increasing values of $t_x$ and $t_y$ simultaneously, we can clearly see the degeneracy breakdown around $t_x=t_y=0.3$. However, this picture is not symmetric and increasing the value of $t_y$ lifts the degeneracy much quicker than the increase of $t_x$. The explanation is that the model is effectively infinite along $OX$ and, as our perturbation theory calculation suggests, in this case, only the terms proportional to $t_y$ can lift the degeneracy coming from the corner spin degrees of freedom as the perturbation theory contributions proportional to $t_x$ are exponentially suppressed by the size of the system. 

\begin{figure}[h]
 \begin{tikzpicture}
    \node[anchor=south west,inner sep=0] at (0,0) {\includegraphics[width=0.45\textwidth]{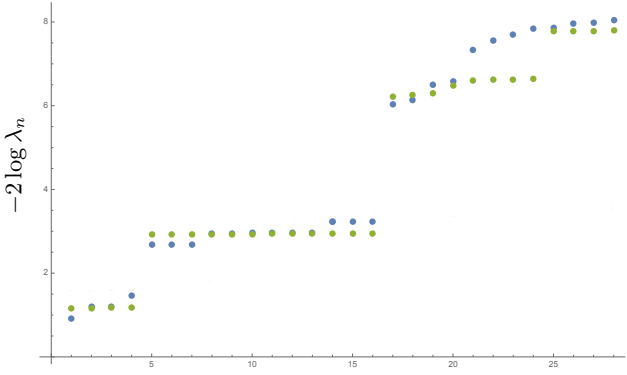}};
    \node [rotate=90] at (-0.2,3.3) {$-2\log\lambda_n$};    
\end{tikzpicture}

	\caption{Entanglement spectrum for the model with $t_x=0.76$, $t_y=0$ computed using two different bond dimensions. Blue dots correspond to $\chi=28$ computation, green dots - to $\chi=48$.}
	\label{spec_28_48}
\end{figure}

This leads to the question why the degeneracy is lifted in the lower right corner of the diagram, where the value of $t_y$ is minimal and whether the degeneracy is restored if we turn off $t_y$ completely. The answer is: there are two factors in play there. First: a slightly non-zero value of $t_y$ and significant values of $t_x$ allow for many different perturbation theory contributions along different paths on the lattice between the corner spins which lift the degeneracy. Second: a fixed bond dimension $\chi=48$ limits the accuracy of our computation. When $t_x$ is increased and the entanglement of the ground state grows, we need to increase the dimension of the MPS 
to accurately simulate the ground state. 
An especially clear example of this phenomena can be seen in the Heisenberg model: if we compute the entanglement spectrum for $t_y=0$ and $t_x=0.76$ using bond dimensions $\chi=28$ and $\chi=48$. This spectrum is shown on Figure \ref{spec_28_48}. 

By eye one can see that the latter computation yields a spectrum which is much closer to being four-fold degenerate. We also show data for exact diagonalization on small systems. The following two plots represent exact diagonalization for the system of two Heisenberg spin chains of length 4, 6, 8, and 10. We can see that as we increase the number of spins, the difference between the entanglement eigenvalues in one quartet falls down exponentially with the factor proportional to $t_y$.


\begin{table}[h]
\begin{ruledtabular}
\begin{tabular}{ c c c c }
\multicolumn{4}{c}{Entanglement spectrum}\\
\hline
 t=0.01 & t=0.1 & t=0.2 & t=0.3\\
\hline
0.4253 & 0.4352 & 0.4680 & 0.5279\\
0.4252 & 0.4274 & 0.4348 & 0.4563\\
0.4252 & 0.4274 & 0.4344 & 0.4562\\
0.4251 & 0.4196 & 0.3983 & 0.3460\\
0.1783 & 0.1770 & 0.1732 & 0.1689\\
0.1783 & 0.1770 & 0.1732 & 0.1680\\
0.1783 & 0.1769 & 0.1720 & 0.1596\\
0.1783 & 0.1768 & 0.1711 & 0.1575\\
0.1783 & 0.1768 & 0.1708 & 0.1596\\
0.1782 & 0.1747 & 0.1633 & 0.1521\\
0.1782 & 0.1747 & 0.1632 & 0.1414\\
0.1782 & 0.1722 & 0.1530 & 0.1411\\
0.0747 & 0.0742 & 0.0725 & 0.1175\\
0.0747 & 0.0735 & 0.0696 & 0.0696\\
0.0747 & 0.0735 & 0.0696 & 0.0627\\
0.0747 & 0.0729 & 0.0668 & 0.0609
\end{tabular}
\end{ruledtabular}
\caption{\label{entsp_table} Entanglement spectrum of an $L_y=4$ quadrupole strip for different values of intra-cell coupling $t$.}
\end{table} 


\begin{figure}[h]
  \begin{tikzpicture}
    \node[anchor=south west,inner sep=0] at (0,0) {\includegraphics[width=0.5\textwidth]{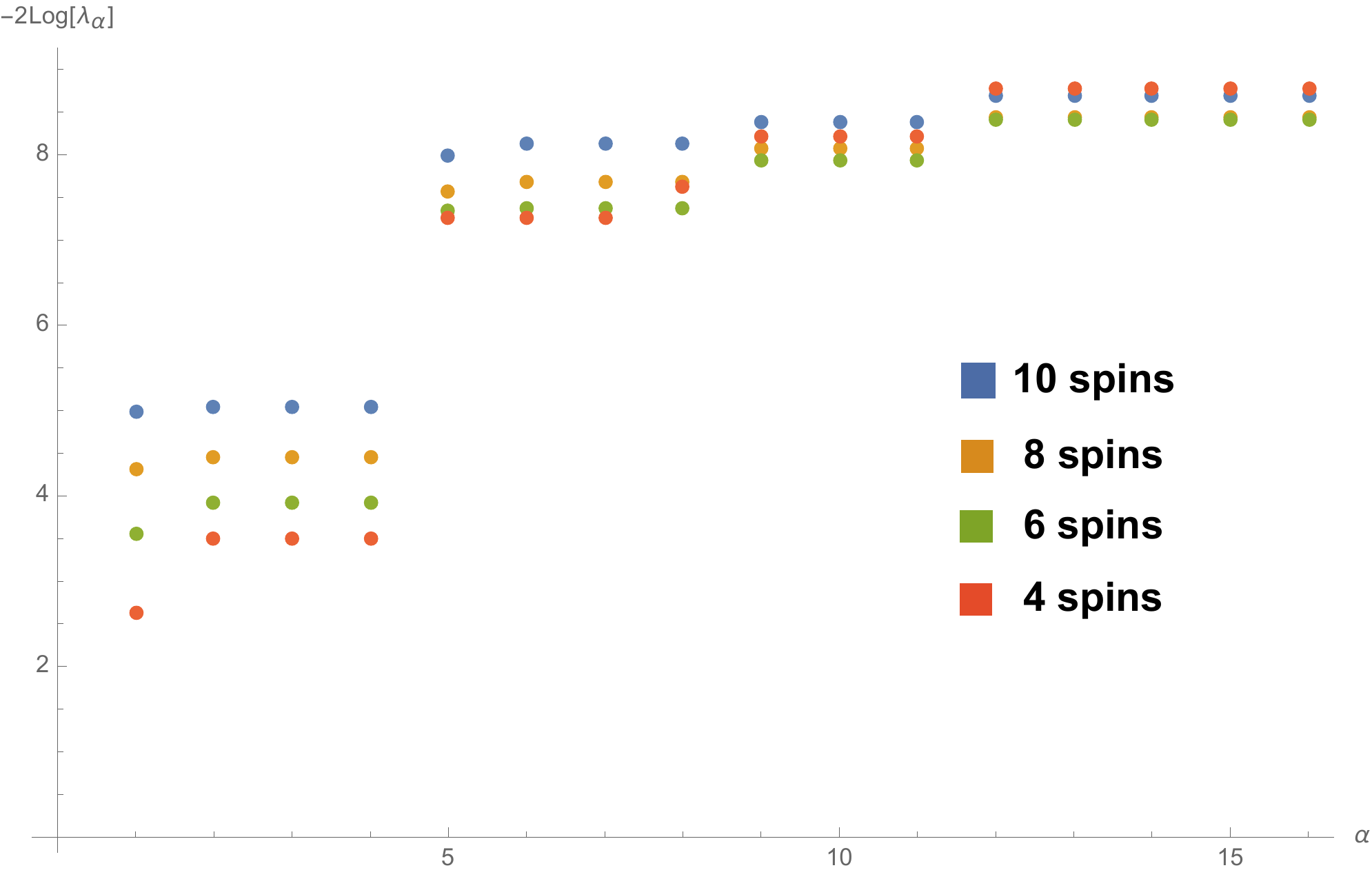}};
    \node at (7.2,3.28) {10 spins};
    \node at (7.2,2.79) {8 spins};
    \node at (7.2,2.32) {6 spins};
    \node at (7.2,1.83) {4 spins}; 
    \node [rotate=90] at (-0.2,3) {$-2\log\lambda_n$};
    \filldraw[fill=white, draw=white] (0,5.4) rectangle (0.8,5.8);
    \filldraw[fill=white, draw=white] (8.7,0) rectangle (9,0.5);
    
\end{tikzpicture}

  \includegraphics[width=0.45\textwidth]{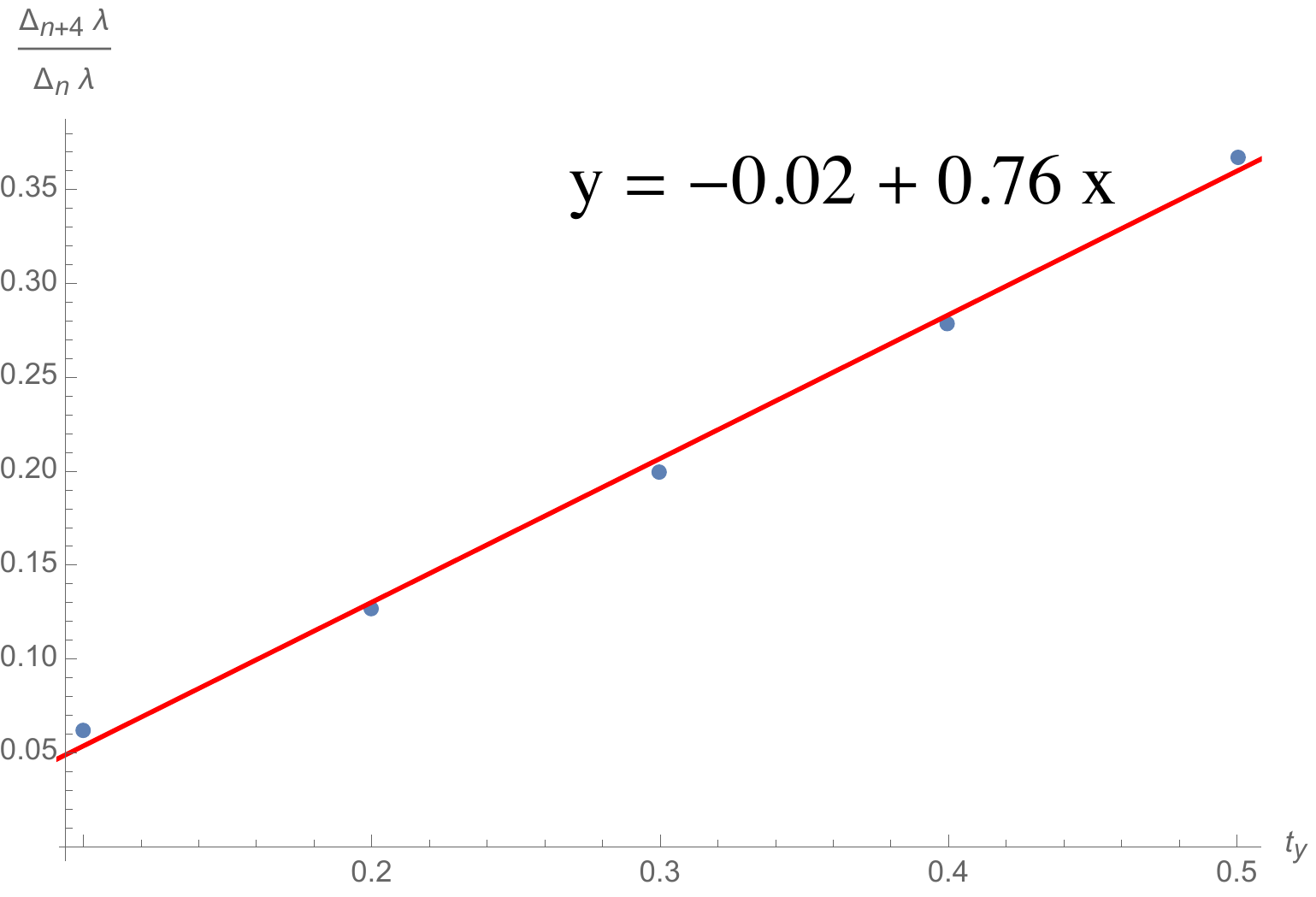}
\end{figure}

\end{document}